\documentclass[article]{revtex4} 
\usepackage{psfig}
\usepackage{bm}
\usepackage{amsmath}
\usepackage{graphicx}

\begin{document}

\title{Data compression and learning in time sequences analysis}

\author{A. Puglisi$^1$, D. Benedetto$^2$, E. Caglioti$^2$, V. Loreto$^1$,
 and A. Vulpiani$^1$}

\affiliation{ $^1$ ``La Sapienza'' University in Rome, Physics
Department, P.le A. Moro 5, 00185 Rome, Italy and INFM, Center for
Statistical Mechanics and Complexity, Rome, Italy\\ $^2$ ``La
Sapienza'' University in Rome , Mathematics Department, P.le A. Moro
2, 00198 Rome Italy}

\date{\today}

\begin{abstract}
Motivated by the problem of the definition of a distance between two
sequences of characters, we investigate the so-called learning process
of typical sequential data compression schemes. We focus on the
problem of how a compression algorithm optimizes its features at the
interface between two different sequences $A$ and $B$ while zipping
the sequence $A+B$ obtained by simply appending $B$ after $A$. We show
the existence of a universal scaling function (the so-called learning
function) which rules the way in which the compression algorithm
learns a sequence $B$ after having compressed a sequence $A$. In
particular it turns out that it exists a crossover length for the
sequence $B$, which depends on the relative entropy between $A$ and
$B$, below which the compression algorithm does not learn the sequence
$B$ (measuring in this way the relative entropy between $A$ and $B$)
and above which it starts learning $B$, i.e. optimizing the
compression using the specific features of $B$. We check the scaling
function on three main classes of systems: Bernoulli schemes,
Markovian sequences and the symbolic dynamic generated by a non
trivial chaotic system (the Lozi map). As a last application of the
method we present the results of a recognition experiment, namely
recognize which dynamical systems produced a given time sequence. We
finally point out the potentiality of these results for segmentation
purposes, i.e. the identification of homogeneous sub-sequences in
heterogeneous sequences (with applications in various fields from
genetic to time-series analysis).
\end{abstract}

\maketitle


\section{Introduction}

The modern approach to time series analysis based on the theory of
dynamical systems and information theory has represented a major
advance in the description and comprehension of a wide range of
phenomena, from geophysics to industrial processes~\cite{kantz_book,
abarbanol}.

Time series represent a particular example of the wider category of
strings of characters which also includes as further examples texts or
genetic sequences (DNA, proteins).  When analyzing a string of
characters the main question is to extract the information it
brings. For example in a DNA sequence this would correspond to the
identification of the subsequences codifying the genes and their
specific functions.

On the other hand for a written text one is interested in {\em
understanding} it, i.e. recognize the language in which the text is
written, its author, the subject treated and eventually the historical
background. More in general one would be interested in the extraction
of specific features and trends that could help in characterizing the
sequences. With a slight misuse of the word we shall refer in the
following to this kind of information as to {\em semantic} information
because it refers to some underlying meaning whose quantification is
far from being a trivial task~\cite{badii}.

In the spirit of having specific tools for the measurements of the
amount of information brought by a sequence, it is rather natural to
approach the problem from a very interesting point of view: that of
information theory ~\cite{shannon,zurek}. Born in the context of
electric communications, information theory has acquired, since the
seminal paper of Shannon~\cite{shannon}, a leading role in many other
fields as computer science, cryptography, biology and
physics~\cite{zurek}.  In this context the word information acquires a
very precise meaning, namely that of the entropy of the string, a
measure of the {\em surprise} the source emitting the sequences can
reserve to us. Besides the notion of entropy, information theory
provides us with a series of tools for more sophisticated measures of
complexity. In this perspective another important concept is related
to the definition of a suitable measure of remoteness between pairs of
sequences. This kind of measure can be crucial for the implementation
of algorithms aimed at recognition purposes.

As it is evident the word information is used with different meanings
in different contexts.  Suppose now for a while to be able to measure
the entropy of a given sequence. It is possible to obtain from this
measure the information (in the semantic sense) we were trying to
extract from the sequence? The answer to this question is again far
from being trivial. We shall nevertheless show in the following that
at least partial answers can be drawn within an {\em a posteriori}
reasoning.

Having this plan in mind the first logical step is to provide us with
some tools to be used in the measurements of the amount of information
contained in a given string or in the relative amount of information
between pairs of strings. In this spirit this paper is mainly focused
on the so-called compression algorithms, i.e. the zippers.

It is well known that compression algorithms provide a powerful tool
for the measure of the entropy and more in general for the estimation
of more sophisticated measures of complexity~\cite{shannon,K57}. Since
the entropy of a string fixes the minimum number of bits one should
use to reproduce it, it is intuitive that a typical zipper, besides
trying to reduce the space occupied on a memory storage device, can be
considered as an entropy meter. Better will be the compression
algorithm, closer will be the length of the zipped file to the minimal
entropic limit and better will be the estimate of the entropy provided
by the zipper.

A great improvement in the field of data compression has been
represented by the so-called Lempel and Ziv 77 algorithm
(LZ77)~\cite{LZ77}. As we shall see in the following, this algorithm
zips a file by exploting the existence of repeated sub-sequences of
characters in it. Its compression efficiency becomes optimal as the
length of the file goes to infinity.

Starting from this idea Merhav and Ziv~\cite{merhav-ziv} proposed a
general method which allows to compute very precisely the relative
entropy between two sequences of characters. A similar approach has
been proposed by Milosavljevi\'c~\cite{Milo}.  Another contribution in
this direction is due to Farach et al.~\cite{farach} in which it is
shown that the best way to estimate the entropy of a source is to
evaluate the relative entropy of two sequences emitted by the source
itself. 

The relative entropy can be considered as a measure of the remoteness
of two sequences and it represents a very important tool to be used
for instance for classification purposes. A severe limitation arises
from the fact that the relative entropy is not a distance in a mathematical
sense: it is not symmetrical and it does not satisfy the triangular
inequality. In many application, like for instance phylogenesis, it is
fundamental to define a real distance between sequences.

In this perspective a very important contribution has been given by
the group of Li in~\cite{li} where it has been proposed a
rigorous definition of distance between unaligned sequences using the
information theoretical concepts of Kolmogorov
complexity~\cite{LiVit}. The computation of this distance is
implemented by means of original data compression
techniques~\cite{gencompress}.

Moreover in the field of the so-called computational linguistics there
have been several contributions showing how data compression
techniques could be useful in solving different problems (an
incomplete list would include
~\cite{bell,teahan,juola,el-yaniv,thaper,khmelev,li_ling,bcl}):
language recognition, authorship recognition or attribution, language
classification, classification of large corpora by subject, etc.  But
of course the possibilities of data-compression based methods go
beyond computational linguistics. Another important example is that of
genetic motivated problems.  Here also there have been important
contributions: for an incomplete bibliography we refer
to~\cite{li,grumbach,Loewenstern} and references therein.  Another
important field of application is represented by the theory of
Dynamical Systems~\cite{benci_1,benci_2,boffetta}.

It is evident how the specific features of data compression techniques
make them potentially very important for fields where the human
intuition can fail: DNA and protein sequences as already mentioned but
more in general geological time series, stock market data, medical
monitoring, etc.

Some of us have recently proposed a method~\cite{bcl} for context
recognition and context classification of strings of characters or
other equivalent coded information.  The remoteness between two
sequences $A$ and $B$ was estimated by zipping a sequence $A+B$
obtained by appending the sequence $B$ after the sequence $A$ (using
the {\em gzip} compressor~\cite{gzip}).  This idea is used for
authorship attribution and, defining a suitable distance between
sequences, for languages phylogenesis.

The idea of appending two files and zip the resulting file in order to
measure the remoteness between them had been previously proposed by
Loewenstern et al.~\cite{Loewenstern} (using {\em zdiff} routines) who
applied it to the analysis of DNA sequences, and by
Khmelev~\cite{khmelev} who applied the method to authorship
attribution. In particular here the method is extensively tested using
many different zippers, including {\em gzip}. Though the idea is the
same the practical implementation differs from the one proposed
in~\cite{bcl}.

On the same stream line other very similar approaches have been
proposed. In particular Teahan (see~\cite{teahan} and references
therein) performs authorship attribution and text classification using
Prediction by Partial Matching algorithms ({\em PPM})~\cite{PPM} data
compression algorithms; Juola (see~\cite{juola} and references
therein) using an algorithm proposed by Wyner to measure relative
entropy perform languages phylogenesis; Thaper~\cite{thaper}
performs experiments of authorship attribution using {\em PPM} and
LZ78~\cite{LZ78} algorithms.

In this paper we extend the analysis of~\cite{bcl} by considering more
in details the features of data compression algorithms when applied to
generic strings of characters.  The specific question we raise here is
how LZ77-like compression algorithms behave at the interface between
two different files. More specifically we shall focus on the process
by which a typical zipper {\em learns} the sequence it is processing
and how it uses previous information acquired while zipping a given
file to zip a second different file. We point out in particular the
existence of a scaling function which rules the way in which the
compression algorithm learns the sequence $B$ after having zipped
sequence $A$. Let us notice how this kind of problems is closely
related to the so-called segmentation problem, i.e. the identification
of homogeneous sub-sequences in heterogeneous sequences (with
applications in various fields from genetic to time-series analysis).

Since in this case we are interested in exploring the features of the
compression algorithms we shall use as benchmark systems time
sequences issued by dynamical systems of increasing complexity. In
particular the scaling function is checked numerically for three main
classes of systems: Bernoulli schemes, Markovian sequences and the non
trivial symbolic dynamic generated by the so-called Lozi map. As a
last application of the method we present the results of a recognition
experiment, namely recognize which dynamical systems produced a given
time sequence.

The outline of the paper is as follows. In section II we recall some
basic definitions. In section III we recall the definition of relative
entropy and the Merhav and Ziv algorithm. In section IV we study
analytically what happens when applying LZ77 algorithm to a sequence
obtained appending two different sequences. In Section V we analyze
numerically the results of section IV. In Section VI we perform a
recognition experiment on sequences generated by the Lozi map. Finally
in section VII we draw the conclusions and discuss possible fields of
application for these techniques.


\section{Basic concepts}
\label{definitions}

Originally information theory (IT) was introduced by
Shannon~\cite{shannon} in the practical context of electric
communications. The powerful concepts and techniques of IT allow for a
systematic study of sources emitting sequences of discrete symbols
(e.g. binary digit sequences) and in the last decades it has been
shown the deep relations between IT and other fields as computer
science, cryptography, biology and chaotic
systems~\cite{zurek,boffetta}.

Consider a symbolic sequence $S(1), S(2), S(3), \dotsc$ where $S(t)$
is the symbol emitted at time $t$ and each $S$ can assume one of $m$
different values. Assuming that the sequence is stationary
we introduce the $N$-block entropy:
\begin{equation}
H_N=-\sum_{\{C_N \}} P(C_N) \ln P(C_N)
\label{h_definition}
\end{equation}

\noindent where $P(C_N)$ is the probability of the $N$-word
$C_N=[S(t),S(t+1),\dotsc,S(t+N-1)]$. The difference

\begin{equation}
h_N=H_{N+1}-H_N
\end{equation}

\noindent has a rather obvious meaning: it is the average information
supplied by the $(N+1)$-th symbol, provided the $N$ previous ones are
known. Noting that the knowledge of a longer past history cannot
increase the uncertainty on the next outcome, one has that $h_N$
cannot increase with $N$ i.e. $h_{N+1} \leq h_N$. Now we are ready to
introduce the Shannon entropy for an ergodic stationary process:

\begin{equation}
h=\underset{N \to \infty}{\lim} h_N = \underset{N \to \infty}{\lim}
\frac{H_N}{N}.
\end{equation}

It is easy to see that for a $k$-th order Markov process (i.e. such
that the conditional probability to have a given symbol only depends
on the last $k$ symbols,
$P(S(t)|S(t-1),S(t-2),\dotsc)=P(S(t)|S(t-1),S(t-2),\dotsc,S(t-k))$)
$h_N=h$ for $N \geq k$.

The Shannon entropy $h$ is a measure of the ``surprise'' the source
emitting the sequence reserves to us.  Let's try to explain in which
sense entropy can be considered as a measure of a surprise.  Suppose
that the surprise one feels upon learning that an event E has occurred
depends only on the probability of E. If the event occurs with
probability 1 (sure!) our surprise in its occurring will be zero. On
the other hand if the probability of occurrence of the event E is
quite small our surprise will be proportionally large. For a single
event occurring with probability $p$ the surprise is proportional to
$\ln P$.  Let's consider now a random variable $X$, which can take $N$
possible values $x_1,...,x_N$ with probabilities $p_1,...,p_N$, the
expected amount of surprise we shall receive upon learning the value
of $X$ is given precisely by the entropy of the source emitting the
random variable $X$, i.e.  $- \sum p_i \ln p_i$.

A theorem, due to Shannon and McMillan~\cite{shannon,K57}, expresses
in a precise way how $h$ quantifies the ``complexity'' of the source:
if $N$ is large enough, the set of $N$-words $\{C_N\}$ can be
partitioned in two classes, $\Omega_1(N)$ and $\Omega_2(N)$ such that
all the words $C_N
\in \Omega_1(N)$ have  probability $P(C_N) \sim \exp(-h N)$
and

\begin{subequations}
\begin{align}
\underset{C_N \in \Omega_1(N)}{\sum} P(C_N) &\to 1 \hspace{1cm}
\text{for} \;\; N \to \infty\\ \underset{C_N \in \Omega_2(N)}{\sum}
P(C_N) &\to 0 \hspace{1cm} \text{for} \;\; N \to \infty.
\end{align}
\end{subequations}

\noindent An important implication of the theorem is that the number
of typical sequences $\mathcal{N}_{eff}(N)$ (those in $\Omega_1(N)$)
effectively observable is

\begin{equation}
\mathcal{N}_{eff}(N) \sim e^{h N}.
\label{number}
\end{equation}

\noindent Note that in non-trivial cases, in which $h < \ln m$,
$\mathcal{N}_{eff}(N) \ll m^N$, being $m^N$ the total number of
possible $N$-words. Let us remark that the Shannon-McMillan theorem
for processes without memory is nothing but the law of large
numbers. Equation~\eqref{number} is somehow the equivalent in IT of the
Boltzmann relation in statistical thermodynamics $S \propto \ln W$,
$W$ being the number of possible microscopic states and $S$ the
thermodynamic entropy. It is
remarkable the fact that, under rather natural assumptions, the
Shannon entropy $h$ apart from a multiplicative factor, is the
unique quantity which characterizes the ``surprise''~\cite{K57}.

An important result is the relation between the maximum compression
rate of a sequence $(S(1),S(2),S(3),\dotsc)$, expressed in an alphabet
with $m$ symbols, and $h$. If the length $T$ of the sequence is
large enough, then it is not possible to compress it into another
sequence (with an alphabet with $M$ symbols) whose size is smaller
than $T h/\ln M$. Therefore, noting that the number of bits needed
for a symbol in an alphabet with $M$ symbol is $\ln M$, one has that
the maximum allowed compression rate is $h/\ln M$. Perhaps the
simplest way to compress, at least at a conceptual level, is via the
Shannon-Fano procedure which is able to reach asymptotically the
maximum allowed compression rate~\cite{welsh}. Also the popular
Lempel-Ziv coding~\cite{LZ77} (see in the following for a short
discussion) gives the same asymptotic results.

We stress the fact that $h$ is an asymptotic quantity which gives
the behavior of $H_N$ (or equivalently $h_N$) at large $N$, i.e. $h
\simeq \frac{H_N}{N}$ for $N \gg 1$. On the other hand the features of
$H_N$ (or $h_N$) for moderate $N$ are rather important in all
nontrivial processes (i.e. with
memory). Grassberger~\cite{grassberger} proposed a way to characterize
the behavior of $H_N$. Let us introduce

\begin{equation}
\delta h_N = h_{N-1}-h_N,
\end{equation}
and the effective measure complexity $C$:

\begin{equation}
C=\sum_{N=1}^{\infty} N \delta h_N.
\end{equation}

It is not difficult to realize that, for large $N$, one has

\begin{equation}
H_N \simeq C+h N.
\end{equation}

In trivial processes (e.g. Bernoulli schemes), $C=0$, on the other
hand $C$ can be nonzero in cases with zero $h$ (e.g. periodic
sequences).

\vspace{0.5cm}

\section{Data compression and Complexity}

As already mentioned it exists an important relation between the
maximum compression rate achievable for a given sequence and its
Shannon entropy. The problem of the optimal coding for a text (or an
image or any other kind of information) has then to face with the
intrinsic limit to encode a given sequence: the entropy of the
sequence.  We have also mentioned that there are many equivalent
definitions of entropy but probably the best definition for our
purposes in this paper is the Chaitin - Kolmogorov
complexity~\cite{K65,Ch66,Ch90,S64}: the algorithmic complexity of a
string of characters is given by the length (in bits) of the smallest
program which produces as output the string.  A string is said complex
if its complexity is proportional to its length. This definition is
really abstract, in particular it is impossible, even in principle, to
find such a program~\cite{LiVit}.  Since this definition tells nothing
about the time the best program should take to reproduce the sequence,
one can never be sure that somewhere else it does not exist another
shorter program that will eventually produce the string as output in a
larger (eventually infinite) time.

Despite this difficulty one has to recall that there are algorithms
explicitly conceived to approach the theoretical limit of the optimal
coding.  These are the file compressors or zippers.  A zipper takes a
file and try to transform it in the shortest possible file.  Obviously
this is not the best way to encode the file but it represents a good
approximation of it.  One of the first compression algorithms is the
Lempel and Ziv algorithm (LZ77)~\cite{LZ77} (used for instance by
$gzip$ and $zip$). It is interesting to briefly recall how it works.
The LZ77 algorithm finds duplicated strings in the input data.  The
second occurrence of a string is replaced by a pointer to the previous
string given by two numbers: a distance, representing how far back
into the window the sequence starts, and a length, representing the
number of characters for which the sequence is identical.  For
example, in the compression of an English text, an occurrence of the
sequence ``the'' will be represented by the pair $(d,3)$, where $d$ is
the distance between the occurrences we are considering and the
previous one.  It is important to mention as the zipper does not
recognize the word ``the'' as a dictionary word but only as a specific
sequence of characters without any reference to the words belonging to
a specific dictionary.  The sequence will be then encoded with a
number of bits equal to $(\log_{2}¥(d)+\log_{2}¥(3) )$: i.e. the
number of bits necessary to encode $d$ and $3$. Roughly speaking the
average distance between two consecutive ``the'' in an English text is
of the order of $10$ characters. Therefore the sequence ``the'' will
be encoded with less then $1$ byte instead of $3$ bytes.

LZ77 algorithm has the following remarkable property: if it encodes
a text of length $L$ emitted by an ergodic source whose entropy per
character is $h$, then the length of the zipped file divided by the
length of the original file tends to $h$ when the length of the text
tends to $\infty$. In other words it does not encode the file in the
best way but it does it better and better as the length of the file
increases.  
More precisely the code rate, i.e. the number of bits per character 
needed to encode the sequence, can be written as:

\begin{equation}
\mbox{code rate} = \frac{\mbox{number of bits to encode the
phrase}}{\mbox{length of a phrase} \times \ln 2}\simeq \frac{\ln n + \ln L_n +
O(\ln \ln L_n)}{L_n \ln 2},
\end{equation}
\noindent where $L_n$ is the length of the phrase substituted and $n$
is the length of the part of the sequence already analyzed. Note that  $\ln n /\ln 2$
is the number of bits needed to encode the part of the pointer
describing the distance, while $\ln L_n / \ln2$ is the number of bits needed
to encode the part of the pointer describing the length of the
substitution. Recalling~\cite{wyner-ziv} that for $n \rightarrow
\infty$ one has that $L_n \rightarrow \frac{\ln n}{h}$ ({\em in
probability}) one obtains

\begin{equation}
\mbox{code rate} \simeq \frac{h}{\ln 2} + \mathcal{O}\left(\frac{\ln \ln n }{\ln n}\right),
\end{equation}
\noindent i.e. the $LZ77$ algorithm converges asymptotically to the
Shannon entropy even though the convergence is extremely slow.  The
presence of the term $\ln 2$ is due to the fact that in the definition
of $h$ in~\eqref{h_definition} we use the natural logarithm.

It is important to remind that the redundancy of the LZ77 coding has
been rigorously determined by Savari in \cite{savari}.

A variant of LZ77 has been introduced in 1978 under the denomination
of LZ78~\cite{LZ78}. In this case the algorithm starts reading the
text. Whenever it meets a new sequence of characters it associates a
new character equivalent to this sequence.  From now onward whenever
it meets this sequence this will be encoded with a single character.
Essentially after a while the algorithm will be able to encode the
typical sequence of characters in a short way.  For LZ78 thus, after a
while, the sequence ``the'' will be then associated to a new
character. More precisely, after a while, the zipper is able to encode
the word ``the'' not with $3$ bytes but only with $3-4$ bits.

The first conclusion one can draw is therefore about the possibility
to measure the entropy of a large enough sequence simply by zipping
it.  For example if one compresses an English text the length of the
zipped file is typically of the order of one fourth of the length of
the initial file. An English file is encoded with $1$ byte ($8$ bits)
per character.  This means that after the compression the file is
encoded with about $2$ bits per character. Obviously this is not yet
optimal. Shannon with an ingenious experiment showed that the entropy
of the English text is something between $0.6$ and $1.3$ bits per
character~\cite{pierce}.

\subsection{Relative entropy}

Another important quantity we need to recall is the notion of relative
entropy or Kullback-Leibler divergence~\cite{KL,Kullback,cover-thomas}
which is a measure of the statistical remoteness between two
distributions.  Its essence can be easily grasped with the following
example.  Let us consider two ergodic sources $A$ and $B$
emitting sequences of $0$ and $1$: $A$ emits a $0$ with
probability $p$ and $1$ with probability $1-p$ while $B$ emits
$0$ with probability $q$ and $1$ with probability $1-q$.  As already
described, the compression algorithm applied to a sequence emitted by
$A$ will be able to encode the sequence almost optimally,
i.e. coding a $0$ with $-\log_2 p$ bits and a $1$ with $-\log_2(1-p)$
bits. This optimal coding will not be the optimal one for the sequence
emitted by $B$.  In particular the entropy per character of the
sequence emitted by $B$ in the coding optimal for $A$ will
be $-q \,\ln p - (1-q) \, \ln (1-p)$ while the entropy per
character of the sequence emitted by $B$ in its optimal coding is
$-q \, \ln q - (1-q) \, \ln (1-q)$. The number of bits per
character wasted to encode the sequence emitted by $B$ with the
coding optimal for $A$ is the relative entropy of $A$ and
$B$,

\begin{equation}
D(B\vert \vert A) \equiv D(q \vert \vert p) = -q \, \ln \frac{p}{q} -
(1-q) \, \ln \frac{1-p}{1-q}.
\end{equation}

A linguistic example will help to clarify the situation: transmitting
an Italian text with a Morse code optimized for English will result in
the need of transmitting an extra number of bits with respect to
another coding optimized for Italian: the difference is a measure of
the relative entropy.

The general definition of the relative entropy for two generic
distributions $p_N({\bf x})$ and $q_N({\bf x})$,
i.e. the probability distributions for sequences ${\bf x}$ of $N$
characters emitted by two different sources, is
given by

\begin{equation}
\label{kullback}
D_N(q_N \vert \vert p_N) = \sum_{{\bf x}} q_N({\bf x}) \ln
\frac{q_N({\bf x})} {p_N({\bf x})}.
\end{equation}

Let us stress that in general $D_N(q_N \vert \vert p_N)$ could be
infinite simply because of sequences emitted by the first source and
not existing in the second (i.e. $p_N({\bf x})=0$ for some sequences
${\bf x}$). 

\subsection{Merhav-Ziv algorithm for the computation of the 
relative entropy}

It is interesting to recall the algorithm proposed by Merhav and
Ziv~\cite{merhav-ziv} for the measurement of the relative entropy.
The method is based on a procedure very similar to the one used in the
LZ77.  This procedure is called parsing and it consists in a
sequential search for phrases each one is the shortest string which is
not a previously parsed phrase (notice that this is a self-parsing
procedure since the algorithm only considers one sequence). In LZ77
this would be the longest string for which there is a match in the
window already analyzed.  The Merhav-Ziv algorithm modifies this
self-parsing procedure by introducing a cross-parsing, i.e. a
sequential parsing of a sequence ${\bf z}$ with respect to another
sequence ${\bf x}$.  In practice one has to find the largest integer
$m$ such that the sub-sequence
$(z_1,...,z_m)=(x_i,x_{i+1},...,x_{i+m-1})$ for some $i$.  The string
$(z_1,...,z_m)$ is defined as the first phrase of ${\bf z}$ with
respect to ${\bf x}$. Next one starts from $z_{m+1}$ and find, in a
similar manner the longest sub-sequence $(z_{m+1},...,z_{n})$ which
appears in ${\bf x}$, and so on. The procedure ends once the entire
sequence ${\bf z}$ has been parsed with respect to ${\bf x}$.  If
$c({\bf z}|{\bf x})$ is the number of phrases in ${\bf z}$ with
respect to ${\bf x}$, Merhav and Ziv demonstrate that for two
sequences of length $n$ the quantity

\begin{equation}
\Delta({\bf z}\|{\bf x}) = \frac{1}{n} [c({\bf z}|{\bf x}) \ln n -
c({\bf z}) \ln c({\bf z})  ],
\end{equation}
\noindent converges as $n \rightarrow \infty$ to the relative entropy
between the two sources that emitted the two sequences ${\bf z}$ and
${\bf x}$. Notice that $c({\bf z}) \ln c({\bf z})$ is the measure of
the complexity of the sequence ${\bf z}$ obtained by a self-parsing
procedure like the one defined by $LZ78$.

It is important to remark how the convergence of the Merhav-Ziv method
to the relative entropy is extraordinarily fast if compared with the
one of LZ77 algorithm for instance, since here the convergence is,
as Merhav and Ziv write~\cite{merhav-ziv}, {\em almost exponential}.
This fact has the important consequence that one can estimate the
complexity of a given sequence more effectively by computing its
relative entropy with respect to itself, more that in an absolute way.


\section{Relative entropy and learning} \label{learning}

In this section we describe how the LZ77 algorithm zips a file obtained
by appending a file $B$ of length $L_{B}$ to a file $A$ of length
$L_{A}$. The files $A$ and $B$ are emitted by two ergodic sources with
ergodic measures given by $p_{A}$ and $p_{B}$ respectively. We will
use the symbols $A$ and $B$ to denote indifferently the files and
their sources.

In particular it is important to understand how the second file is
encoded once the sequential zipper starts reading it.  Very roughly
what happens is the following.  First of all the zipper encodes file
$A$. Then it begins encoding file $B$. Initially the zipper will find
the longest match of the file $B$ in the file $A$. After a while
however, longer is the fraction of $B$ already analyzed, larger will
be the probability to find the longest match in file $B$ itself.
Asymptotically the longest matches of file $B$ will be found only
inside $B$. This means that we can roughly describe this process as a
two step process: in a first time the zipper tends to optimize the
coding for the $A$ part while in a second time it encodes the $B$ file
with the coding obtained for the $A$ part (transient) as well as with
the statistics proper of the $B$ file (which will asymptotically
dominate). For these reasons the zipping procedure of $A+B$ can be
seen as a sort of learning process.

It is convenient here to consider the following idealized problem.
Let $\sigma$ be an infinite sequence extracted with measure $p_{B}$.
Let $\sigma_{A}$ a sequence of length $L_{A}$ extracted with the
measure $p_{A},$ and $\sigma_{B}$ a sequence of length $L_{B}$
extracted from the measure $p_{B}$. Let us define the function
$P(L_{A},L_{B})$ as the probability that the longest subsequence of
$\sigma_A$ matching a subsequence of $\sigma$ is longer than the
longest subsequence of $\sigma_B$ matching a subsequence of
$\sigma$. In other words $P(L_A,L_B)$ is the probability that $\sigma$
finds a longer part of itself in $\sigma_A$ rather than in $\sigma_B$.
Moreover let us define $\tilde{P}(\ln L_{A},\ln
L_{B})=P(L_{A},L_{B})$.

Once fixed the notation we can go back to the original problem. In
this case $P(L_A,L_B)$ will be the probability that, once the zipper
starts scanning the $B$ part of the $A+B$ file, it finds a
matching in the $A$ part rather than in the $B$ part.

We can say that the typical distance between two occurrences of the
same substring is inversely proportional to the probability of the
substring itself.  Denoting with $d(B \vert \vert A)$ the relative
entropy (for symbols), an argument based on the Shannon-McMillan
theorem~\cite{merhav-ziv} shows that the probability of occurrence of
a string of length $n$ of the sequence $\sigma$ with respect to the
measure $p_{A}$ is asymptotically given by $e^{-n[h_B+d(B\vert \vert
A)]}$. We are assuming that $d(B \vert \vert A)$ is a well defined
finite limit of \eqref{kullback} for large $N$ (we shall see in the
following that this is not a serious limitation for the experiment):

\begin{equation}
d(B \vert \vert A)=\lim_{N \to \infty} \frac{1}{N} d(q_N \vert \vert
p_N).
\end{equation} 

Therefore the length $n_{A}$ of the longest match found in $A$ will be
approximately obtained by imposing $L_{A} e^{-n[h_B+d(B\vert \vert
A)]}=1$, whose inversion gives

\begin{equation}
n_{A}={\ln L_A\over h_B+d(B\vert\vert A)}.
\end{equation}

\noindent Analogously the length of the longest match found in the
part of file $B$ already encoded will be given approximately by
\begin{equation}
n_{B}={\ln L_{B} \over h_B}.
\end{equation} 

\noindent Therefore we expect that if
\begin{equation}
{\ln L_{B} \over h_B}<<
{{\ln L_A}\over {h_B+d(B \vert \vert A)}}
\end{equation} 
\noindent the longest match will be found in $A$,
i.e. $P(L_{A},L_{B})\simeq 1$, while if

\begin{equation}{
\ln l\over h_B}>>{\ln{L_A}\over h_B+d(B \vert \vert A)}
\end{equation} 
\noindent one expects to find it in $B$, i.e. $P(L_{A},L_{B})\simeq
0$.

It is important now to focus more precisely on the transient region
where, as already noticed, it takes place a sort of learning process.
In order to do this we first consider the case in which the two
sequences $A$ and $B$ are $(0,1)$ Bernoulli sequences of length
$L_{A}$ and $L_{B}$ respectively. Afterward we shall try to generalize
the result.

\noindent The first source emits $0$ with probability $p$ and $1$ with
probability $1-p.$ The second source emits emits $0$ with probability
$q$ and $1$ with probability $1-q$. Therefore, in a typical sequence
of length $n$ emitted by the second source, $0$ will appear
approximately $q n$ times while $1$ will appear approximately $(1-q)n$
times.  More precisely we can say that $m_{0}$ (the number of zeros in
the second sequence) is approximately a Gaussian random variable with
average $q n$ and variance $O(n)$.

By neglecting the fluctuations of $m_{0}$ one has that the probability
of this sequence with respect to the measure of the first source will
be approximately given by

\begin{equation}
p^{q n} (1-p)^{(1-q)n} = e^{n\left[q\ln p+(1-q)\ln
(1-p)\right]}.
\end{equation}

\noindent This expression is nothing that $e^{-n[h_B+d(B\vert \vert
A)]}$, confirming, at least in this particular case, the previous
result.

Now let us take into account the fluctuations. $m_{0}$ has random
fluctuations of order $\sqrt{n}$ around its average. This fluctuations
induces fluctuations of the probability of this string with respect to
the measure $p_{B}$. We then expect fluctuations of order
$n=O(\sqrt{\ln L})$ of the length $n_{A}$ of the longest match found
in the first string. The same is true for $n_{B}.$ This leads us to
conjecture that, as $L_{A},L_{B}\rightarrow \infty,$ $\tilde{P}(x,y)$
converges to a unique function under this scaling, i.e.

\begin{equation}
\tilde{P}(x,y)\rightarrow_{x,y\rightarrow\infty} f({{x}-\alpha y\over
{\sqrt{x+y}}}),
\label{scaling}
\end{equation}

\noindent where $\alpha={h_B\over h_B+ d(B \vert \vert A)}.$ On the
basis of large deviations theory~\cite{varadhan}, we expect this
conjecture to be valid for sequences with short term memory,
i.e. where the correlations decay sufficiently fast. In the next
section we shall numerically check this conjecture.

The rigorous analysis of the fluctuations of the length of the longest
match is a very interesting and difficult problem (see~\cite{aldous}
and~\cite{jacquet}).

\section{Numerical results}

The hypothesis for the scaling form $\tilde{P}(x,y)=f\left(
\frac{x-\alpha y}{\sqrt{x+y}}\right)$ introduced in the previous
section for the so-called learning function, can be tested for finite
size symbol sequences generated according to some stochastic rule,
e.g. with pseudo-random number generators or with some kind of non
trivial dynamical systems. In this section we shall check this
hypothesis in three cases featuring an increasing {\em complexity}:
simple Bernoulli schemes, Markov processes and Non-Markovian processes
(using in particular the symbolic dynamics generated by the Lozi map).

\subsection{Bernoulli scheme}

The simplest random sequence of symbols is generated by a Bernoulli
scheme: at each time $t$ the symbol $S(t)$ is $0$ with probability $p$
and $1$ with probability $1-p$, with $p \in [0,1]$. This is the
sequence of biased (unfair) coin tosses. In this case it is very easy
to see that $h=h_N=\frac{H_N}{N}=-[p\ln p+(1-p)\ln (1-p)]$ for every
$N \geq 1$, and the effective measure complexity is $C=0$.

We have generated a sequence $A$ of $0$'s and $1$'s of length $L_A$
using a Bernoulli scheme with a probability $p_A$ for $0$'s, and then
a set of $5000$ sequences $B$ of length $L_B^{max}$ where $0$'s occur
with probability $p_B$. For these cases the relative entropy (for
sequences of any length) is given by

\begin{equation}
d(p_B \vert \vert p_A) = p_B\ln(p_B/p_A) + (1-p_B)\ln((1-p_B)/(1-p_A))
\end{equation}

For each sequence of this set, the following
numerical experiment has been performed:

\begin{enumerate}

\item
A sequence $AB$ (of length $L_A+L_B$) is obtained appending the $B$
sequence to the end of the $A$ sequence;

\item
One starts scanning the sequence $AB$ from the point
$i=i_{start}=L_A+1$, i.e. from the first character of the sequence $B$;

\item  \label{start}
One looks for the longest sub-sequence that:

\begin{enumerate}

\item
starts at $i$ and, 

\item
is identical to a sub-sequence contained in the part $[1,i]$ of the joint sequence $AB$.

\end{enumerate}

\noindent The length of this maximum sub-sequence is called $n_{max}$.

\item
The index $i$ is increased by $n_{max}$. If $i<L_A+L_B^{max}$ the algorithm
goes to~\ref{start}, otherwise the algorithm stops.

\end{enumerate}

In the above procedure, one keeps track of the statistics of the
sub-sequence matchings: in particular we are interested to the number
of sub-sequences found in $A$ or in $B$ as a function of
$L_B=i-i_{start}$. At the beginning of the scanning procedure most of
the matchings are found in $A$. When $L_B$ is large enough,
sub-sequence matchings found in $B$ can be competitive with their
length against the ones found in $A$. The procedure of averaging over
many ``realizations'' of sequence $B$ allows us for a smooth
statistics, i.e. a smooth curve $P(L_A,L_B)$ vs. $L_B$ with fixed
$L_A$.

\begin{figure}[h]
\begin{center}
\includegraphics[width=9cm,keepaspectratio,clip=true]{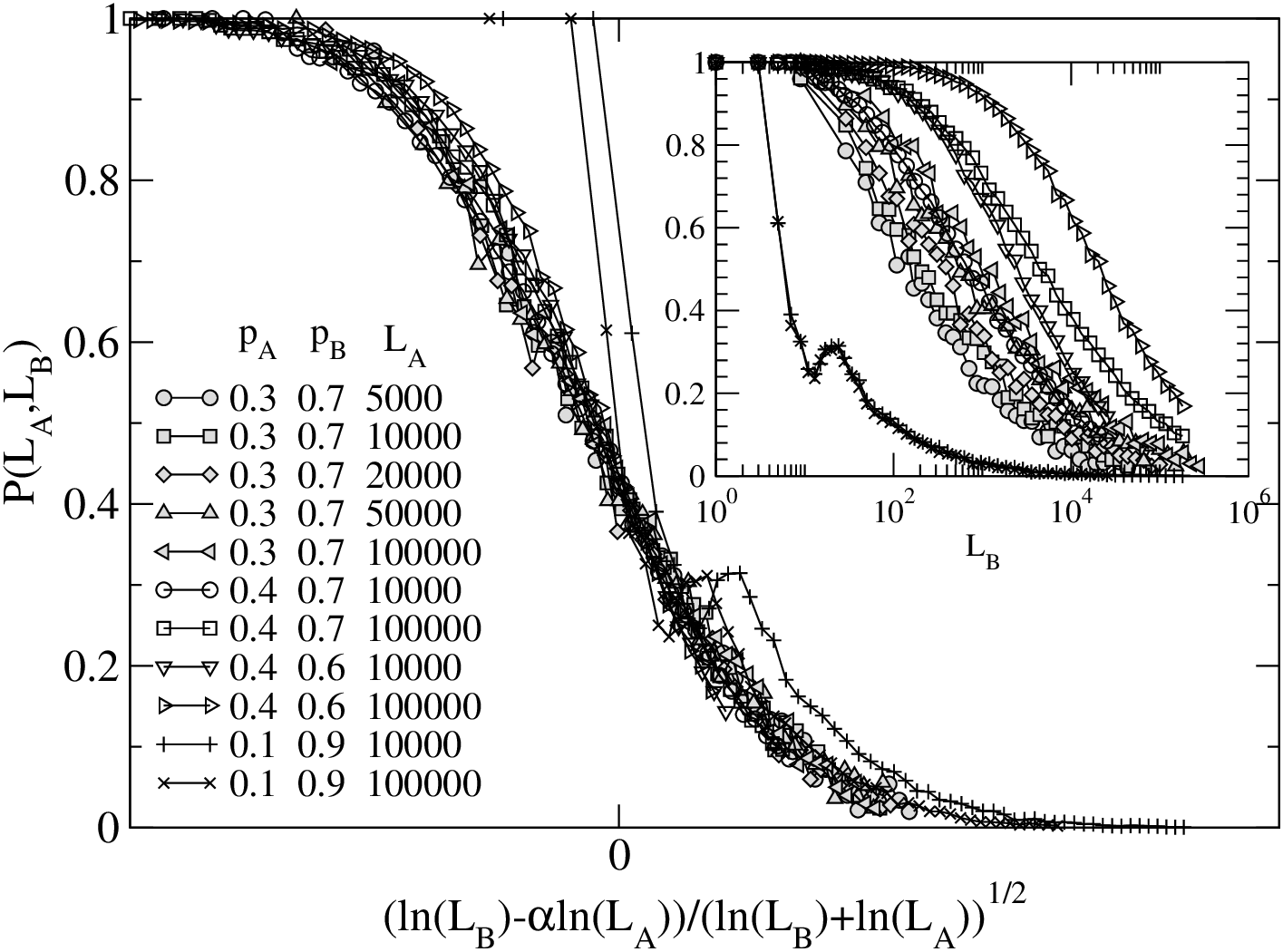}
\end{center}
\caption{ Collapse of $P(L_A,L_B)$ versus the rescaled coordinate
(discussed in the text), for different pairs of Bernoulli processes
with different probabilities of symbol ``zero'' $(p_A,p_B)$ and with
different lengths $L_A$ of buffer $A$. In the inset the same data are
shown versus $L_B$, i.e. without any rescaling. The values of $\alpha$
are the following: 0.643 for $(p_A,p_B)=(0.3,0.7)$, 0.768 for
$(p_A,p_B)=(0.4,0.7)$, 0.892 for $(p_A,p_B)=(0.4,0.6)$, 0.156 for
$(p_A,p_B)=(0.1,0.9)$. 
\label{bernoulli_scaling}}
\end{figure}

Fig.~\ref{bernoulli_scaling} reports the curves obtained with the
above procedure for $P(L_A,L_B)$ vs. $L_B$ for different values of
$L_A$ and different choices of the pair $(p_A,p_B)$, as well as their
collapse using the scaling function \eqref{scaling}. The collapse is
indeed very satisfying, bringing the first evidence for the conjecture
in \eqref{scaling}. In the picture is also shown the failure of the
scaling form when $\alpha$ is too small ({\it pluses} and {\it
crosses} in the inset, not reported in the main plot). This happens
when the two sequences are too different or when the second sequence
has an entropy $h$ very low: in both cases the convenience of parsing
the sub-sequences of $B$ with sub-sequences of its own past (and not
from $A$) comes too early, as can be seen in the inset of the
figure. As a consequence of this, the length of $A$ does not matter
for the parsing of sequence $B$ and the two curves obtained with
different $L_A$ (those with $\alpha=0.156$) are identical. For those
curves, the scaling obviously fails.

\subsection{Markovian sequences}

The logical step after having tested the scaling conjecture
(\ref{scaling}) on Bernoulli schemes, is a test using sequences
generated by means of Markov chains. A Markov chain is a random
process with discrete space of states, where the probability of every
state is determined by one or more previous states. A Markov chain is
therefore a Bernoulli process generalized to any number of symbols and
provided with memory. The order of Markov chains is the number of
previous states influencing the present: e.g. for a Markov chain of
order $k=1$ the probability of having a certain symbol depends only on
the previous symbol and is determined by its conditional probability
$W_{ij}=P(S(t)=j|S(t-1)=i)$. We have tested the scaling hypothesis on
the Lempel-Ziv parsing procedure of pairs of two symbols, order one,
symmetric Markov chains. This means that both $A$ and $B$ are
sequences of $0$'s and $1$'s and that their transition matrix is of
the form:

\begin{equation}
W=
\begin{pmatrix}
w   &1-w \\
1-w &w
\end{pmatrix}
\label{sym_markov}
\end{equation}

\noindent with $w \in [0,1]$ the probability of repeating the previous
symbol. The sequences $A$ and $B$ have different transition matrices,
that is $w=w_A$ for $A$ and $w=w_B$ for $B$. In practice a sequence
obtained with $w$ near $1$ is something like $11111100000011111100000
\dotsc$, while a sequence obtained with $w$ near $0$ is like
$010101001010101101010\dotsc$.

For a long sequence of this kind the probability $P(S)$ is stationary
and $H_N=H_1+(N-1)h$, that is $C=H_1-h$. Moreover we are interested in
the quantity (in the following referred to as {\it block cross
entropy}) $\tilde{h}_N=\tilde{H}_{N+1}-\tilde{H}_N$ vs. $N$, where

\begin{equation}
\tilde{H}_N=-\sum_{\{C_N \}^*} P_B(C_N) \ln P_A(C_N)=H^B_N+D_N(B \vert \vert A)
\end{equation}

\noindent with $P_A(C_N)$ the frequency of the $N$-sequence $C_N$ in
the sequence $A$ (and analogously for $P_B(C_N)$) and $\{C_N \}^*$ is
the set of $N$-sequences contained both in $A$ and in $B$. If we
consider infinite sequences $A$ and $B$ and a two state Markov process
(as the one introduced in this section) then $\{C_N \}^* \equiv \{C_N
\}$, i.e. the whole set of $2^N$ sequences of length $N$ is explored
by both dynamics. For the kind of Markov chain described by the
transition matrix in \eqref{sym_markov}, we can therefore calculate

\begin{subequations}
\begin{align}
\tilde{H}_N &= \tilde{H}_1-(N-1)\sum_{\{S_i S_j \}} P_B(S_i) W^B_{ij}
\ln W^A_{ij} \\ \tilde{h}_N &= - \sum_{\{S_i S_j \}} P_B(S_i) W^B_{ij}
\ln W^A_{ij}= h_A+d(B \vert \vert A).
\end{align}
\end{subequations}

\noindent 
More in general the above formulas hold if $W^A_{ij}$ is positive when
$W^B_{ij}$ is positive.

\begin{figure}[h]
\begin{center}
\includegraphics[width=9cm,keepaspectratio,clip=true]{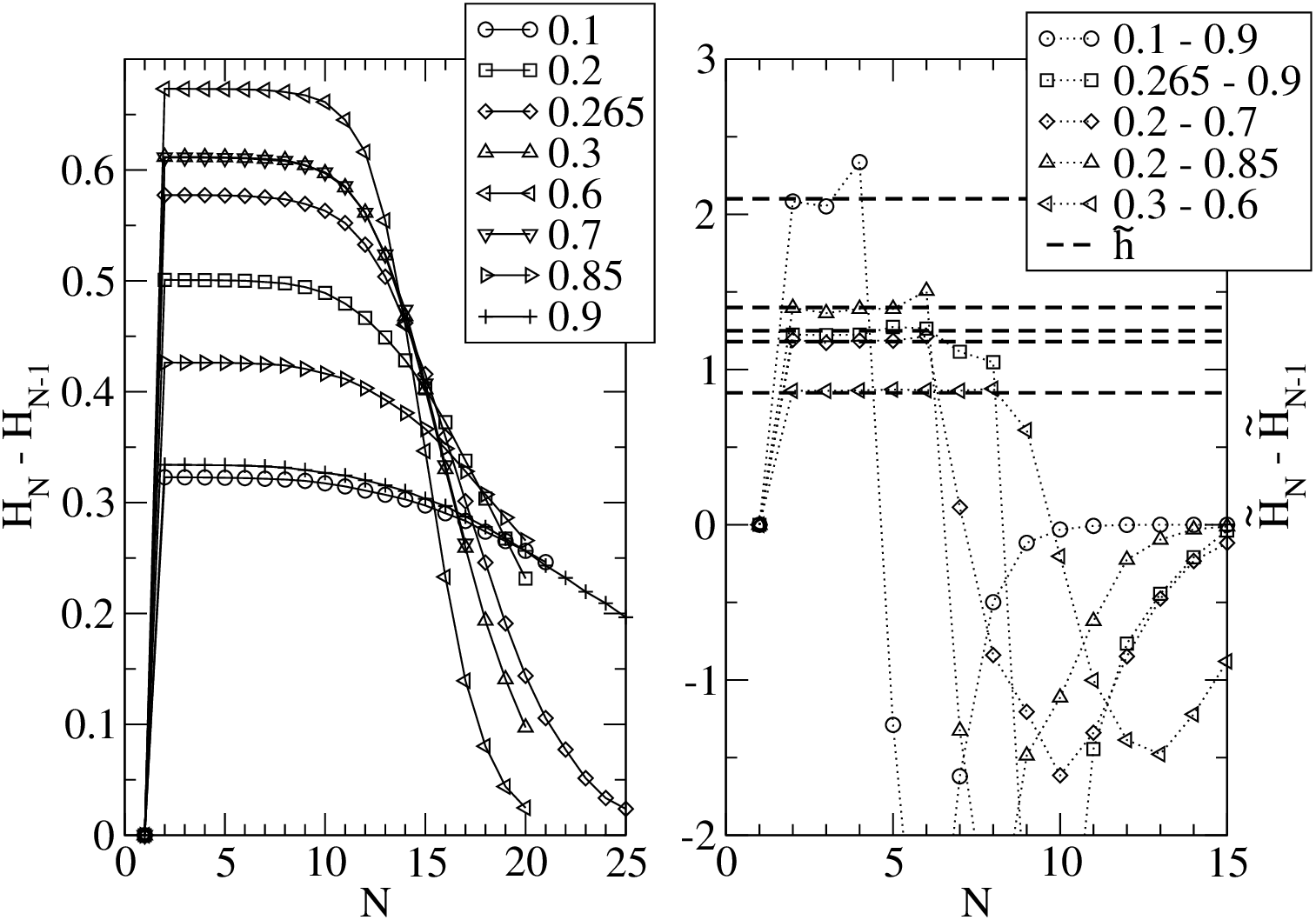}
\end{center}
\caption{ Left: $H_N-H_{N-1}$ vs. $N$ for a Markov process of order
$1$ with symmetrical transition matrix (see eq.~\eqref{sym_markov})
calculated numerically using a sequence of $20000$ symbols, for
different values of the parameter $w$. The plateau (reached at $N=2$)
corresponds to the theoretical $h$, while the successive decay of the
curves is due to poor statistics. Right: cross entropy
$\tilde{h}_N=\tilde{H}_N-\tilde{H}_{N-1}$ for different pairs $(A,B)$
of such Markov processes, characterized by parameters $w_A,w_B$. The
plateau (put in evidence by dashed lines) correspond to the
theoretical value $\tilde{h}$.
\label{markov_entropy}}
\end{figure}

In fig.~\ref{markov_entropy} we show the effects of finiteness of the
sequences $A$ and $B$ on the block entropy $h_N$ and on the block
cross entropy $\tilde{h}_N$: for finite sequences $A$ and $B$, even in
the case of two state Markov chains, the sets of words of length $N$
may not coincide. $A$ and $B$ are sequences of length $20000$
generated with the symmetric one-step Markov processes with different
transition matrices $W$, i.e. with different parameters $w_A$ and
$w_B$.

\begin{figure}[h]
\begin{center}
\includegraphics[width=9cm,keepaspectratio,clip=true]{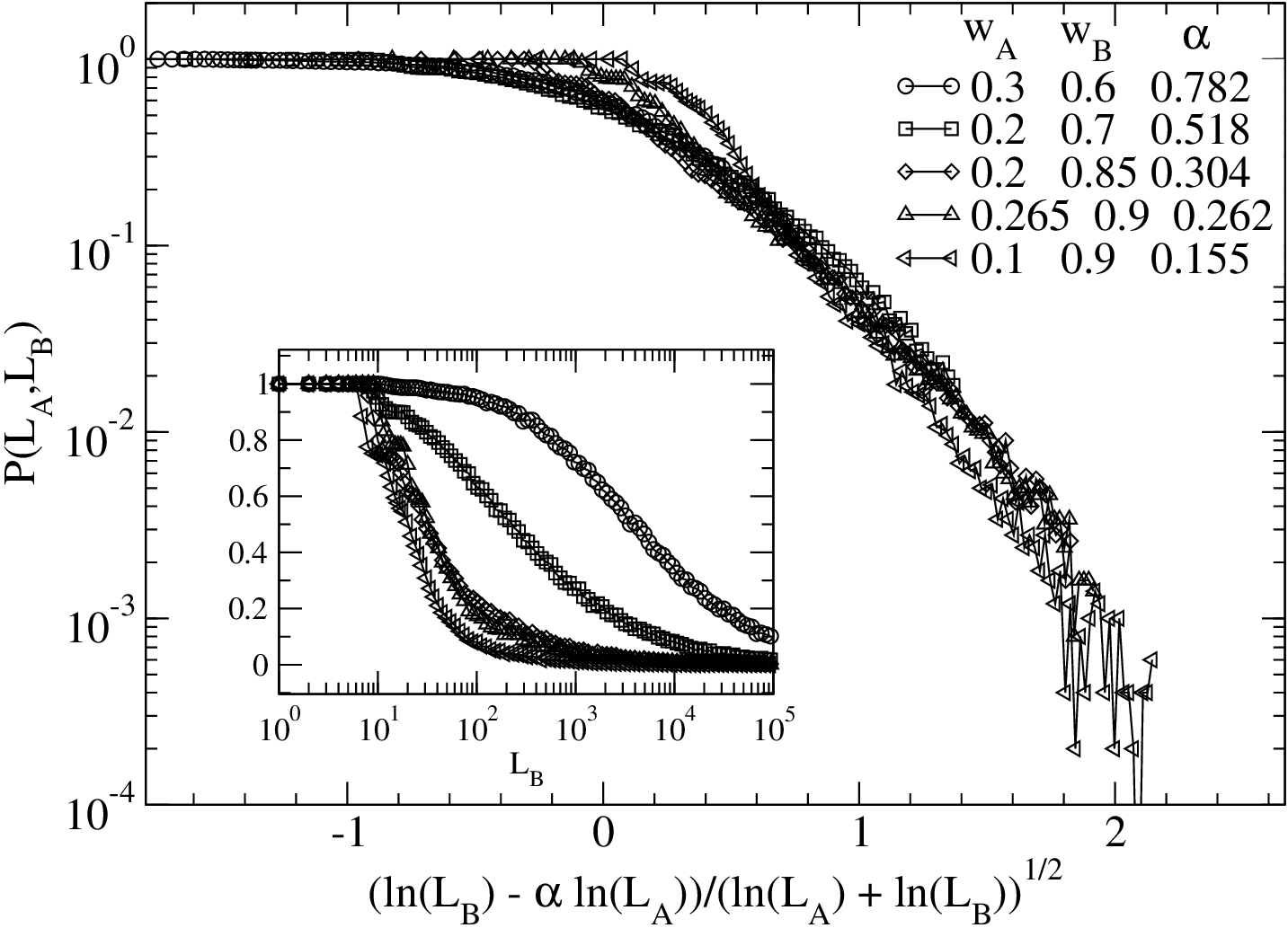}
\end{center}
\caption{ Collapse of $P(L_A,L_B)$ versus the rescaled coordinate for
Markov processes of order $1$ and symmetric transition matrix (see
eq.~\eqref{sym_markov}) with different values of the pairs
$(w_A,w_B)$, with $L_A=20000$. In the figure are also indicated the
values of $\alpha$, see~\eqref{scaling}. In the inset the data without
rescaling.
\label{markov_scaling}}
\end{figure}

It can be seen that the plateau representing $h$ is reached at $N=2$,
as expected for Markov chains of order $k=1$. Moreover, it can be seen
the effect of finite size: the sequences considered are $20000$
symbols long, therefore, invoking the Shannon-McMillan theorem, one
has that $N$ must be not too large in order to satisfy the condition
that the number of typical $N$-sequences be much smaller than the
length of the sequence, i.e. $\mathcal{N}=\exp(hN) \ll
20000$. Otherwise the statistics becomes too poor and $h_N$ rapidly
departs from $h$. In the right plot of fig.~\ref{markov_entropy} we
show the behaviour of $\tilde{h}_N$: the first plateau of the curves
in this graph provides an estimate of the ``cross entropy''
$\tilde{h}=d(B\vert \vert A)+h_B$, where $d(B \vert \vert A)$ is the
Kullback-Leibler entropy relative to the pair of sequences $A$ and
$B$, while $h_B$ is the Shannon entropy of sequence $B$. This figures
shows how finite size effects appear in the computation of $d(B \vert
\vert A)$, well before those appearing in the computation of $h$; this
is a direct consequence of the operative definition used in this
computation: in order to have a good estimate of $\tilde{h}_N$ a large
amount of $N$-sequences common both to $A$ and $B$ is indeed needed,
reducing the value of the finite size cut-off. The scaling of
$P(L_A,L_B)$ for pairs of Markov sequences is shown in
figure~\ref{markov_scaling}. Again a good collapse is obtained using
the previously proposed scaling form~\ref{scaling}. It is also clear
that the collapse fails for pairs of processes with $\alpha \ll 1$,
i.e. the pairs with the strongest difference in the transition matrix.

\subsection{Non-Markovian sequences: Lozi map symbolic dynamics}

It is interesting to probe a class of signals (i.e. sequences) with a
higher degree of complexity than simple Markov chains. Dynamical
systems are a rather natural source of non-trivial signals. A symbolic
sequence can be associated to the dynamical system by means of a
partition of the phase space $\Omega$, i.e. $\{ \omega_i \}$ with $m$
elements such that $\bigcup_{i=1}^m \omega_i = \Omega$ and $\omega_i
\cap \omega_j=0$ for every $i$ and $j$ in $[1,m]$. Every trajectory
$\mathbf{x}(t)$ is therefore mapped into a sequence of symbols whose
values are the $m$ symbols of the alphabet. An interesting non-trivial
example is the symbolic dynamics obtained with a binary partition of
the $x$ variable of the Lozi map, defined as:

\begin{subequations}
\begin{align}
x(n+1) &= -a|x(n)|+y(n)+1 \\
y(n+1) &= bx(n)
\end{align}
\end{subequations}
where $a$ and $b$ are parameters. The sequence of symbols used in the
following test is obtained taking $0$ when $x \le 0$ and $1$ when $x >
0$. For $b=0.5$, numerical studies show that the Lozi map is chaotic
for $a$ in the interval $(1.51,1.7)$. For a discussion of the Lozi
map, computation of Lyapunov exponents and representation of its
symbolic dynamics in terms of Markov chains, see~\cite{book_vulpio}.

\begin{figure}[h]
\begin{center}
\includegraphics[width=9cm,keepaspectratio,clip=true]{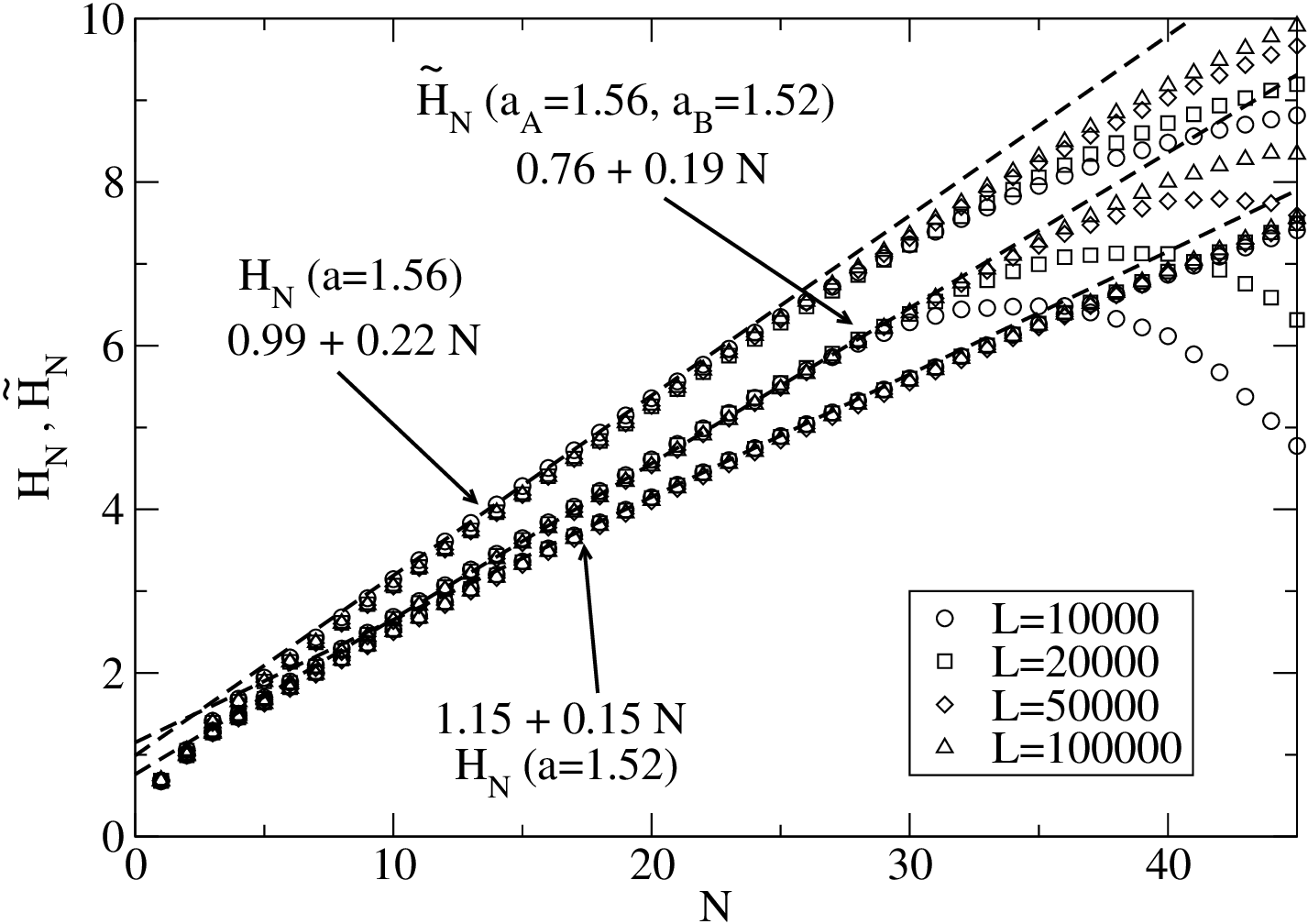}
\end{center}
\caption{ $H_N$ and $\tilde{H}_N$ versus $N$ for sequences of symbols
obtained with a binary partition of the Lozi map. The $H_N$ are
calculated using Lozi map with parameter $a=1.52$ and $a=1.56$. The
$\tilde{H}_N$ are calculated using pairs of Lozi map with $a_A=1.56$
and $a_B=1.52$. All calculations have been performed with sequences of
different length $L$, to probe finite size effects.
\label{lozi_entropy2}}
\end{figure}

Fig.~\ref{lozi_entropy2} reports the numerical computation of $H_N$
and $\tilde{H}_N$ (the relative block cross entropies) for several
sequence lengths, using always the same pair of processes $a_A=1.56$
and $a_B=1.52$. The aim is putting in evidence finite size effects as
well as estimating Shannon and Kullback-Leibler entropies needed for
the collapse of $P(L_A,L_B)$. The estimate of $d(B \vert \vert A)$ and
$h_B$ and therefore of $\alpha$ is obtained with a level of confidence
of $10\%$.

More specifically Fig.~\ref{lozi_entropy2} is particularly
enlightening from the point of view of the meaning of the effective
measure complexity $C$. A naive order $1$ Markovian approximation of
the map is far from reproducing the dynamical properties of the Lozi
map. This can be appreciated in Fig.~\ref{lozi_entropy2}, noting that
$C$ is not small.

\begin{figure}[h]
\begin{center}
\includegraphics[width=9cm,keepaspectratio,clip=true]{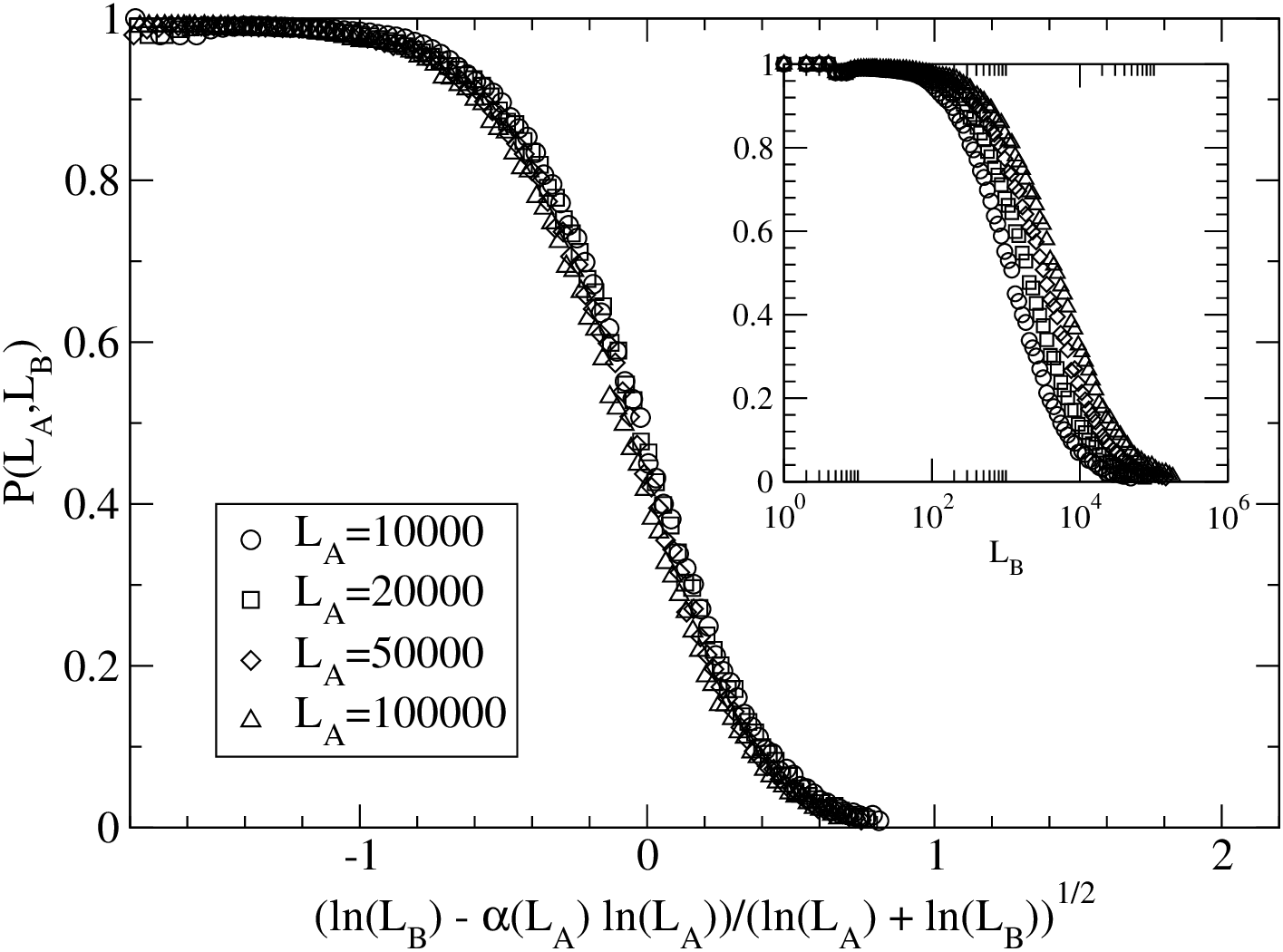}
\end{center}
\caption{ Collapse of $P(L_A,L_B)$ versus the rescaled coordinate for
sequences of symbols obtained with a binary partition of the Lozi map
with parameters pairs $(a_A,a_B)=(1.56,1.52)$, using an estimate of
$\alpha=0.78$ obtained using the values $h_B=0.15$ and $\tilde{h}=0.19$ (see
Fig.~\ref{lozi_entropy2}). In the inset the same data are shown versus
$L_B$, i.e. without rescaling.
\label{lozi_scaling}}
\end{figure}

Finally, in Fig.~\ref{lozi_scaling} it is shown that the collapse of
the learning curves $p(x,y)$ is very well verified, using again
averages on the $B$ sequence (i.e. different initial conditions) and
different lengths for the $A$ sequence.

Let us note that in the symbolic sequence generated by the Lozi map,
for a given value of the parameter $a$, some words are forbidden and,
in general one has different forbidden sequences for different values
of $a$. Therefore the limit $\lim_{N \to \infty} \frac{1}{N} D_N(B
\vert \vert A)$ does not exists. Nevertheless using as $d(B \vert
\vert A)$ the value extrapolated from moderate values of $N$ one has a
nice agreement with the theoretical argument.

\section{An experiment of recognition}

The last set of results concerns one of the main motivation of this
analysis, i.e. its practical applications. The algorithm proposed in
\cite{khmelev,bcl} has its main justification in its efficiency on the
framework of sequence recognition: the algorithm is able to provide an
estimate of the Kullback-Leibler entropy of a sequence of unknown
provenance relatively to a set of sequences whose provenance is
certain (known sources) and used as reference sequences, giving the
most ``similar'' sequence and therefore the most probable source for
the sequence of unknown provenance. In this context, we have checked
that this recipe well recognizes a symbolic sequence drawn from the
class of Lozi maps. Though the results are very preliminary and a
systematic analysis should be in order, some interesting conclusions
can be drawn.

\begin{figure}[h]
\begin{center}
\includegraphics[width=9cm,keepaspectratio,clip=true]{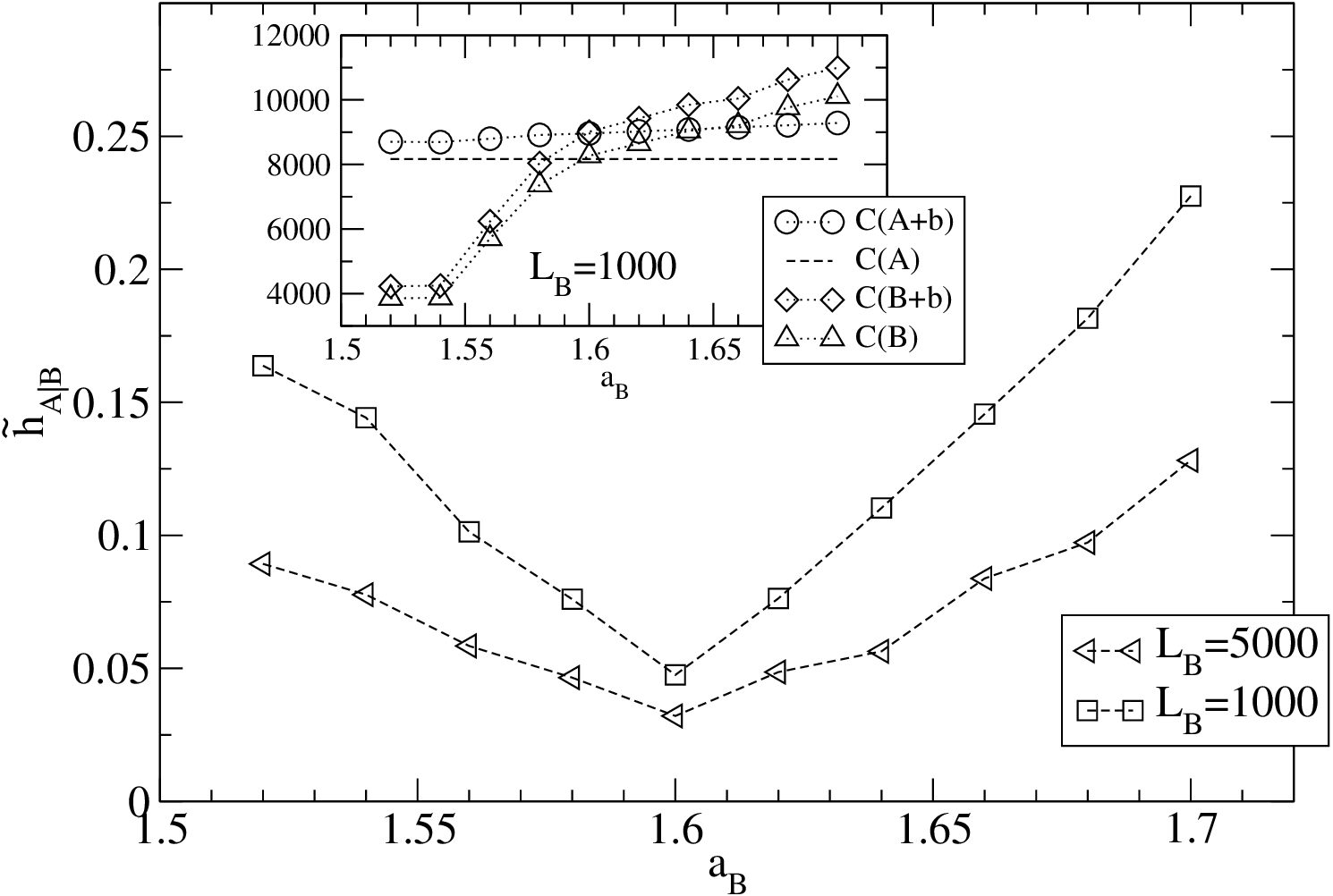}
\end{center}
\caption{ Estimate, by means of LZ77 compression, of $D(B \vert \vert
A)$ (see text) of the Kullback-Leibler entropy relative to different
pairs $(A,B)$ of sequences of symbols: each pair is composed by a
fixed sequence $A$ obtained as a binary partition of a Lozi map with
parameter $a_A=1.6$ and a variable sequence $B$ obtained either as a
binary partition of a Lozi map with variable parameter $a_B$. The
sequences have length $L_A=L_B=10000$. The estimate of the
Kullback-Leibler entropy has its minimum in correspondence of the pair
$(A,A)$ (i.e. when $B$ comes from a Lozi map with $a_B=a_A$): this
indicates that this estimate of $D(B\vert \vert A)$ is capable of
recognizing in the space of Lozi maps. In the inset the lengths of the
LZ-compressed sequences are reported, where $B^*$ is always a sequence
of the same kind of $B$ (note that $L_B^\alpha \simeq 1300$ and
therefore $L_B^*=1000$, and $L_B^*=5000$ are below and beyond the
crossover threshold respectively)
\label{lozi_minimum}}
\end{figure}

Fig.~\ref{lozi_minimum} reports the result of this test. A Lozi map
with $a=1.6$, $b=0.5$ and initial condition $x=0.1$, $y=0.1$ has been
used to generate the sequence $A$, of length $10000$, that will be
used as unknown sequence. As probing sequences we have generated two
sets of sequences, $B$ and $B^*$ respectively, obtained with Lozi maps
with the parameters $b=0.5$ and $a_B=a_{B^*}$ varying between $1.52$
and $1.7$. The sequences $B$ has length of $10000$ while sequence
$B^*$ has length of $5000$ or $1000$. The quantities plotted in the
inset are the lengths of the compressed code (with the LZ77 algorithm,
see the discussion in paragraph~\ref{definitions}), i.e. $C(X)$ is the
length of the code obtained by compressing the sequence $X$. Data
relative to the compression of the sequences $B+B^*$ and $A+B^*$ have
been obtained by averaging over $100$ different choices of initial
conditions. The quantity computed and reported in the main graph is an
estimate of the Kullback-Leibler entropy $d(B \vert \vert A)$, as a
difference (per bit) between $C(A+B^*)-C(A)$ and $C(B+B^*)-C(B)$ which
are the estimates of $d(B \vert \vert A)+h_B$ and $h_B$
respectively. The bottom plot shows very well how this simple recipe
leads to a perfect recognition of the correct value of $a=1.6$: the
estimate of the Kullback-Leibler entropy has in fact an absolute
minimum for that value.

In figure~\ref{lozi_minimum} one can also appreciate the usefulness of
the theoretical analysis of section~\ref{learning}, i.e. the fact that
$L_A^\alpha$ is a good estimate of the best length $L_B$ of the probe
sequences $B$ to obtain the optimal resolution in the recognition
process. In fact in section~\ref{learning} we conjectured (and
successively verified with numerical experiments) that when $L_B$ is
smaller than the cross-over length $L_A^\alpha$, the LZ77 algorithm is
encoding the sequence $B$ with the ``language'' of $A$ and therefore
the length of the encoded sequence is effectively a measure of the
distance between the two languages. Using the previous value
$\alpha=0.78$ as a rough estimate for every other choice of the map
parameter $a$, and given $L_A=10000$, one obtains for the crossover
length $\sim 1300$. In the figure, the resolution power of the LZ77
algorithm with $L_B=1000$ is much higher than that with $L_B=5000$.

\section{Conclusions}

We have studied the properties of standard sequential compression
algorithms in the problem of information extraction from sequences of
characters. We have in particular analyzed the learning process that
these algorithm perform when they are used to compress heterogeneous
data, i.e. data coming from different sources. 

The typical benchmark for this study is a finite sequence of $L_A+L_B$
symbols obtained appending a sequence of $L_B$ symbols emitted by a
source $B$ to a sequence of $L_A$ symbols emitted by a source $A$. An
algorithm like {\em LZ77}~\cite{LZ77}, after having processed the $A$
part of the sequence, starts encoding the $B$ part using the
knowledge acquired while zipping the $A$ part; after a transient the
compression algorithm starts encoding the $B$ part using the knowledge
coming only from the $B$ part already processed (i.e. the zipper
starts learning the $B$ part). We have made a scaling hypothesis that
characterizes this transient process in terms of the entropy of the
source $B$ and the Kullback-Leibler divergence between the two
sequences. 

We have studied the finite size scaling (i.e. incorporating
fluctuations due to the finite size of the sequences under
investigation) by means of numerical experiments on three sets of data
coming from different sources: the Bernoulli scheme, the Markov chain
of first order (with symmetric transition matrix) and the symbolic
dynamics obtained with a binary partition of the Lozi map.  These
three examples feature an increasing complexity: the Bernoulli scheme
emits sequences of uncorrelated random symbols; the Markov chain of
first order is the simplest way to enforce correlations among symbols
in the sequences; finally the Lozi map has the property of having an
higher effective measure complexity~\cite{grassberger}. The scaling
hypothesis is very well verified in all the cases investigated,
pointing out the generality of the result.

These results have a practical importance in the analysis of a
recently proposed scheme that computes the informational remoteness
between two sequences~\cite{bcl}: in fact this scheme employs a
variant of the LZ77 algorithm and gives the best estimate of the
remoteness (Kullback-Leibler divergence) when the length of the second
sequence is chosen of the order of the threshold value of the learning
function we have introduced in this work. We have investigated
quantitatively this point, showing that the resolution power of the
recognition scheme proposed in~\cite{bcl} is highly improved when the
length of the second sequence is chosen according to the analysis of
the transient. Sequences too short or too long can give bad estimates
of the Kullback-Leibler divergence and therefore a big uncertainty in
the recognition of similar sequences.

Another important field of application is that of the segmentation of
heterogeneous sequences, i.e. the identification of the boundaries
between regions featuring very different properties which, depending
on the sequences considered, can correspond to very different
phenomena (catastrophic events in geophysical time series, or
boundaries between different sections in genetic sequences just to
quote a couple of examples). In all these cases one could try to
exploit the features of data compression techniques at the interface
between heterogeneous regions in order to define and optimize suitable
observables sensitive to sudden changes.

{\Large Acknowledgements} V. L. and A. P. acknowledge support from the
INFM {\em Center for Statistical Mechanics and Complexity} (SMC).

\end{document}